\documentclass[final,5p,times,authoryear]{elsarticle}

\usepackage{epsfig}

\usepackage{amssymb}

\usepackage{newtxtext,newtxmath}
\usepackage[T1]{fontenc}

\usepackage{subcaption}
\usepackage{csquotes}
\usepackage{multicol}
\usepackage{comment}
\usepackage[dvipsnames]{xcolor}

\usepackage{gensymb}

\usepackage{changepage}

\DeclareMathOperator*{\argmin}{arg\,min}

\usepackage{lastpage}

\def\bsp{\vspace{0.5cm}\normalsize\noindent This paper
has been typeset from a \TeX / \LaTeX\ file prepared by the author.}

\usepackage{, bm}

\usepackage{hyperref}
\usepackage{graphicx}
\usepackage{amsmath}
\usepackage{float}
\usepackage{commath}
\usepackage{geometry}
\usepackage{tabularx}
\usepackage[toc,page]{appendix}
\usepackage[noadjust]{cite}
\usepackage{aas_macros}

\journal{Astronomy and Computing}

\begin{document}

\begin{frontmatter}




\title{\textbf{Reconstructing the mid-infrared spectra of galaxies using ultraviolet to submillimeter photometry and Deep Generative Networks}}
%


\author[NORTH]{
Agapi Rissaki\corref{cor1}}

\cortext[cor1]{Corresponding author:}
\ead{rissaki.a@northeastern.edu}
\author[EUC]{Orestis Pavlou}
\author[NTUA]{Dimitris Fotakis}
\author[EUC]{Vicky Papadopoulou Lesta}
\author[EUC]{Andreas Efstathiou}

\affiliation[NORTH]{
            organization={Khoury College of Computer Sciences, Northeastern University},
            addressline={440 Huntington Avenue, 202 West Village H}, 
            city={Boston},
            postcode={MA 02115}, 
            country={USA}}

\affiliation[EUC]{
            organization={School of Sciences, European University Cyprus},
            addressline={Diogenes Street, 6, Engomi}, 
            city={Nicosia},
            postcode={1516}, 
            country={Cyprus}}
            
 \affiliation[NTUA]{
            organization={National Technical University of Athens},
            addressline={Heroon Polytechniou 9, Zografou}, 
            city={Athens},
            postcode={15780}, 
            country={Greece}}




\begin{abstract}
The mid-infrared spectra of galaxies are rich in features such as the Polycyclic Aromatic Hydrocarbon (PAH) and silicate dust features which give valuable information about the physics of galaxies and their evolution. For example they can provide information about the relative contribution of star formation and accretion from a supermassive black hole to the power output of galaxies. However, the mid-infrared spectra are currently available for a very small fraction of galaxies that have been detected in deep multi-wavelength surveys of the sky. In this paper we explore whether Deep Generative Network methods can be used to reconstruct mid-infrared spectra in the 5-35$\mu m$ range using the limited multi-wavelength photometry in $\sim 20$ bands from the ultraviolet to the submillimeter which is typically available in extragalactic surveys. For this purpose we use simulated spectra computed with a combination of radiative transfer models for starbursts, active galactic nucleus (AGN) tori and host galaxies. We find that our method using Deep Generative Networks, namely Generative Adversarial Networks and Generative Latent Optimization models, can efficiently produce high quality reconstructions of mid-infrared spectra in $\sim$ 60\% of the cases. We discuss how our method can be improved by using more training data, photometric bands, model parameters or by employing other generative networks.
\end{abstract}
 

\end{frontmatter}


\section{Introduction}
\subsection{Motivation}
The spectral energy distributions (SEDs) of galaxies contain important information on the physics of the galaxies and their evolutionary stage. Originally large galaxy surveys like SDSS \citep{Gunn2006} and 2MASS \citep{Skrutskie2006} covered the optical and near-infrared wavelength range. Recently, however, multi-wavelength surveys covering the ultraviolet to millimetre spectrum have been made possible with missions like GALEX, Spitzer, Herschel, WISE (\citet{Lonsdale2003}, \citet{Eales2010}, \citet{Oliver2012}).  Projects like the Herschel Extragalactic Legacy Project (HELP; \citet{Shirley2019,Shirley2021}) created archives with ultraviolet to millimetre SEDs for millions of galaxies.

The study of the panchromatic SEDs of galaxies from the ultraviolet to the millimeter (0.1-1000$\mu m$) is essential for the study of galaxies and their evolution because of the presence of interstellar dust which absorbs about half of the optical and ultraviolet radiation emitted by stars and accreting supermassive black holes and reemits it in the infrared. For a recent discussion of this topic and various methods of studying SEDs see \citet{Perez2021}.

Of particular interest are luminous infrared galaxies (LIRGs with 1-1000$\mu m$ luminosities exceeding $10^{11}L_\odot$), ultraluminous infrared galaxies (ULIRGs with 1-1000$\mu m$ luminosities exceeding $10^{12}L_\odot$ and hyperluminous infrared galaxies (HLIRGs with 1-1000$\mu m$ luminosities exceeding $10^{13}L_\odot$). Analysis of their SEDs can give an indication of whether the dominant power source of their observed luminosity is mainly star formation, an active galactic nucleus (AGN) powered by accretion on their central supermassive black hole or a combination of the two processes \citep{Farrah2003,Vega08,LoFaro2015,Efstathiou2022}. This can be explored with SED fitting methods using radiative transfer models or so-called energy balance methods \citep{Efstathiou2003,DaCunha2008,Boquien2019} to extract physical quantities of interest for the galaxies such as stellar mass, star formation rate, active galactic nucleus (AGN) fraction etc.

In this effort it is particularly useful to have the mid-infrared spectra of the galaxies ($\sim 5-35\mu m$) which are rich in important features such as the Polycyclic Aromatic Hydrocarbon (PAH) emission features and silicate dust absorption/emission features. The mid-infrared spectra of galaxies can also be used to construct diagnostic diagrams for galaxy classification, such as the \textit{fork} diagram introduced by \citet{Spoon2007}. For a few thousand galaxies, 5-35$\mu m$ spectra obtained with the IRS instrument onboard NASA's Spitzer Space Telescope \citet{Houck04} are available (\citet{Lebouteiller2011,Lebouteiller2015,Spoon2022}). However, for most of the millions of galaxies detected in large multi-wavelength surveys of the sky (e.g. \citet{Shirley2019,Shirley2021}), only limited photometry is available in about 20 bands ranging from the ultraviolet to the submillimetre part of the spectrum. 

The recently launched James Webb Space Telescope (JWST) will provide more and better quality mid-infrared spectra than Spitzer but it can only observe up to 28$\mu m$ so it can only provide data for the 9.7$\mu m$ silicate feature for $z < 1.8$.  Following the abandonment of the SPICA mission (e.g. \citet{Spinoglio2017}), there is currently no planned mission which will allow observations of the rest frame mid-infrared spectra of galaxies at $z > 2$.

\subsection{Contribution}

In this paper we aim to use spectra of luminous infrared galaxies to train deep generative models, such as Generative Adversarial Network (GAN) and 
Generative Latent Optimization (GLO) models, in order to reconstruct the mid-infrared spectra of galaxies using only the photometry in about 20 bands from the optical to the submillimeter, which is usually available with projects like HELP. Deep learning methods have an inherent demand for substantial training data, which is unavailable in our case. We overcome this limitation of data scarcity by employing \textit{simulated spectra}. Specifically, for the training and the evaluation of our deep generative models we use simulated spectra computed with the CYGNUS (CYprus models for Galaxies and their NUclear Spectra) radiative transfer models (\citet{efstathiou95,efstathiou00,efstathiou09,Efstathiou2021,Efstathiou2022}). Such models have been used in a number of studies to fit the multiwavelength SEDs of LIRGs and ULIRGs, e.g. \citet{herr17,kool18,matt18,Efstathiou2022,Papaefthymiou2022}.

In a nutshell, we aim to show that carefully designed and adequately trained deep generative models, such as GAN and GLO models, are capable of efficiently reconstructing mid-infrared spectra from very sparse photometry signals so that the reconstructed spectra are essentially indistinguishable from the spectra generated by the CYGNUS radiative transfer models used in previous works (see e.g., \citet{Efstathiou2022} and the references therein). 

The approach we follow can be outlined as follows: We simulate 10,000 spectra of galaxies by considering a combination of emissions from a starburst, an active galactic nucleus (AGN) torus and a spheroidal host galaxy. These simulated spectra have the resolution which is typical of modern radiative transfer models. They have in particular 223 wavelength bins covering the range 0.03$\mu m$ to 12mm. They also cover fairly well the mid-infrared part of the spectrum which as noted above includes the PAH emission features and silicate dust features. We then interpolate these spectra to predict the fluxes in 20 bands from the optical to the submillimeter (see Table 1). We assume that all of the galaxies are at redshift = 2 motivated by the fact that there are currently no planned upcoming missions for observing rest-frame mid-infrared spectra of galaxies at this redshift, as described in Section 1.1. At the same time at z=2 there are sufficient photometry data to allow the reconstruction. Of course, the method can be applied at any redshift provided training data at each redshift are generated and processed.

For these 10,000 simulated galaxies we therefore have both the photometry in 20 bands and the 5-35$\mu m$ mid-infrared spectra. 

We then develop, train and test Generative Adversarial Networks (GANs) (\citet{gans}) that aim to reconstruct the 5-35$\mu m$ spectra from the photometry. We assess how successful this reconstruction is by comparing the reconstructed 5-35$\mu m$ spectra with the corresponding simulated spectra.
Additionally, to evaluate the success of GANs in this task, we utilize the framework of the Generative Latent Optimization (GLO) \citep{glo}, as an alternative training method of the generator with simple reconstruction losses. 
Interestingly, we found that GLOs can achieve significantly better reconstruction of the signal than GANs. We attribute the improved performance of GLO-based models compared against the performance of more popular GANs to the fact that GANs usually require a large sample of training data in order to perform well, while GLOs achieve good reconstruction even with relatively small sets of training data. Since the training data sets of our reconstruction problem are not sufficiently large, GLOs perform better than GANs.  

\subsection{Related Work}
 
Generative Adversarial Networks (GANs) is a game-theoretic (\citet{Nisan2007}) deep generative model that can be trained from a sample data set in order to generate new data samples which are statistically similar to the samples in the training set. 
The novelty of a GAN compared to other generative models is that the training is achieved through a game played between two neural network models, which correspond to the players of the game: the ``\textit{generator}'', which learns to generate new plausible samples, and the ``\textit{discriminator}'', which  learns to distinguish artificially generated samples from real ones. During the game play, the generator (resp. the discriminator) improves in generating (resp. distinguishing real from) artificially generated samples. At the end of the training processes (where a Nash equilibrium of the corresponding two-player game is reached), the discriminator showcases a $\sim 50\%$ success rate in distinguishing real from artificially generated samples, indicating that it can no longer clearly distinguish real from artificial samples produced by the generator.

Deep learning approaches and GANs have been successfully applied to a variety of settings spanning from image restoration, high-fidelity natural image synthesis (\citet{Brock2019}), medical image processing (\citet{Topol2019}), medical informatics, missing value imputation (\citet{Yoon2018}) to cybersecurity, fashion and advertising (see e.g., \citet{Pan2019} for a survey).
 
GANs recently gained considerable attention in the field of astrophysics and cosmology. GANs were successfully applied for generating fast novel 2D and 3D images of the large-scale structure of $\Lambda$CDM model universes (\citet{Mustafa2019,Curtis2020,Ullmo2021}), for recovering features and improving upon the quality of astrophysical images of galaxies (\citet{Schawinski2017,Lauritsen2021}), for high-resolution cosmological simulations (\citet{Yin2021,Rodriguez2018}), and for cosmological and astrophysical simulations (\citet{Zamudio2019,Villaescusa2021,Zamudio2019}). Furthermore, \citet{Tamosiunas2021} explored GANs for producing weak gravitational lensing convergence maps and dark matter overdensity field data for multiple redshifts and data of unseen cosmological parameters.
Interestingly, most of these works explore GANs for the production/improvement/analysis of astrophysical/cosmological images or simulations. Closer to our application of interest, \citet{kalmbach2017estimating} applied traditional machine learning techniques to estimate spectra from photometry. This work illustrated that machine learning can provide an effective toolbox for such problems, however the methodology and results were limited to traditional methods that lack the power of deep learning techniques. To our knowledge, our work is among the first ones that explores deep generative models for problems in astrophysics that are not related to image reconstruction. In particular, our work initiates the application of GANs to reconstructing spectral energy distributions of galaxies. 
This application of GANs and GLOs for the extrapolation and generation of reliable signal reconstructions of galaxy spectra can provide a very useful tool for the identification of signal properties, leading to the classification of galaxies by accurately reconstructing missing parts of their spectra. This approach could prove to be very important for multiwavelength astrophysical and cosmological studies of different types of astrophysical targets, by addressing the problem of obtaining complete, panchromatic observations of astrophysical spectra across a wide range of wavelengths, which is one of the major challenges of modern astronomy. For example, reconstructed mid-infrared spectra can be used to construct fork diagrams and other classification diagrams of LIRGs and ULIRGs, as in \citet{Spoon2007} and \citet{Papaefthymiou2022}, at redshifts that will be inaccessible for the forseeable future. The reconstructed spectra can also be used to carry out the graph theoretical analysis described in \citet{pavlou2023}.
 
\paragraph*{\textbf{Roadmap}}
This paper is organized as follows. In section \ref{theoretical_framework_and_methodology} we describe the theoretical framework as well as our methodology of implementing GANs and GLOs for the reconstruction of galaxy spectra. In section \ref{experimental_setting} we present the experimental data used in our work, the training configuration and our reconstruction procedure. In section \ref{results} we showcase our results. In section \ref{discussion} we discuss our results and finally we present our conclusions in section \ref{conclusions}.

\section{Theoretical Framework and Methodology}\label{theoretical_framework_and_methodology}

In this section we present the  formulation of the problem, the methodology followed and the theoretical framework related to deep generative models.

 \subsection{Deep Generative Models}
 \label{background_section}

In Machine Learning literature, we often come across two types of models: discriminative and generative models. Discriminative models aim to discriminate between data coming from different groups, i.e., by modeling the conditional distribution of a certain input admitting a certain group label. On the other hand, generative models aim to directly model a data distribution, usually in a probabilistic manner.

In the context of deep learning, a deep generative network is a deep learning model that functions as a generative model. Such a network learns to approximate a complex data distribution, usually lying in a high dimensional space, via a sufficiently large sample from that distribution. After successful training, the network is able to generate new samples from the data distribution of interest. 

The input to the network is a low-dimensional vector called \textit{latent vector} or \textit{latent code} and the corresponding input space is called the \textit{latent space}. The core assumption is that the data have some level of redundancy and mapping them to lower dimensions will reveal the structure of the complex data distribution, resulting in \textit{data compression}.The latent vectors cannot be explicitly interpreted, although they typically encode properties which will be reflected in the corresponding output. The latent space may have specific structure, e.g., a manifold, and it can be either random or learned. For each model we use, we will explain these properties in detail in later sections. 

We formally define a deep generative network or \textit{generator} $G(\cdot)$ as a mapping:

$$
G \; : \; \mathcal{Z} \rightarrow \mathcal{X}\,,
$$
from the latent space $\mathcal{Z}$ to the output domain $\mathcal{X}$. A generator $G(\cdot)$ is typically trained on a specific dataset, the training set $\mathcal{X}_{\text{train}} \subset X$, in order to learn the underlying data distribution. For example a deep generative network that generates cat images will be trained using thousand or even millions of images of different cats. After training, one could input a random latent vector to the network and receive as output an image of a cat. In this example, we can imagine that the latent vector could determine the cat's breed or coat color. 

In general, a deep generative model can be thought of as a representation of a data distribution in a certain domain. Thus, it can be used as a structural prior by determining the probability of a given sample under said distribution. In what follows, we will explain the details of how we can create and use a deep generative model as a structural prior, which we will call a \textit{deep generative prior}.

 \subsection{Reconstruction Using a Deep Generative Prior}\label{Reconstruction Methods_section}

To formally define our reconstruction problem, we utilize the (noisy) compressed sensing formulation bellow:
\begin{equation} \label{noisyCS}
y = Ax^* + \eta\,,
\end{equation} 
where we collect (possibly noisy) measurements $y \in \mathbb{R}^{m}$ from an underlying signal $x^* \in \mathbb{R}^{n}$, where $m, n\geq 0$ and $m \ll n$. The measurement matrix $A \in \mathbb{R}^{m \times n}$ describes a linear measurement procedure, which is possibly corrupted by additive noise denoted by the noise vector $\eta \in \mathbb{R}^{m}$. The general goal of solving an inverse problem, as the one above, is to reconstruct the signal $x^*$, given the measurement matrix $A$ and few measurements $y$. We expect the signal to lie in a high dimensional space, whereas the size of measurements is significantly smaller than the ambient dimension, i.e., $m \ll n$. 
Although the above formulation refers to compressed sensing, it can also model several real-world inverse problems, e.g., denoising, where we assume that the signal is corrupted by additive noise, inpainting, where a continuous \enquote{window} of the signal is missing, super-resolution, where a percentage of the signal's dimensions is not measured, etc. 

In the classic compressed sensing literature, signal reconstruction typically involves a structural prior, i.e., certain structural assumptions about the underlying signal. The most popular assumption is sparsity, i.e. few non-zero elements, which allows reconstruction from few measurements by solving an under-determined linear system. In general, a structural prior allows efficient signal reconstruction by reducing the search space of possible signals. This can be achieved by determining which signals display the target characteristics (e.g., sparsity).

In this work, we use deep generative networks as structural priors, a method that was first proposed by \citet{bora17} and has been successfully applied to natural images. In this case, we can reduce our search space to the signals that belong to the data distribution modelled by the generative prior. 

The following sections describe the theoretical framework for signal reconstruction as well as the methods for creating the deep generative models utilized in this work.  

\subsubsection{Reconstruction Methods with a Generative Prior}

Given the generative network $G(\cdot)$, the estimated solution of an inverse problem \eqref{noisyCS} could be $\hat{x} = G(\hat{z})$ where:
\begin{equation} \label{recon1}
\hat{z} = \argmin\limits_{z\in \mathcal{Z}} \frac{1}{m} ||AG(z) - y||^2_2  
\end{equation} 
  where  $m$ is  the number of available measurements for the reconstruction problem. In other words, we (approximately) optimize the latent code $\hat{z}$ such that the corresponding signal $\hat{x}$ causes a measurement $A \hat{x}$ that matches the actual measurement $y$. We optimize $\hat{z}$ by back-propagating the gradient of the reconstruction loss through $G(\cdot)$ \citep{bora17}. 

Most generative networks are trained so that the latent space has some geometric structure, e.g., a manifold. In this case, $z$ may have to be regularized so that it reflects the structure of the latent space of $G(\cdot)$. One option is to project $z$ onto the unit sphere, which explicitly restricts the considered latent space and is useful for models discussed in Section \ref{glo_training}.
In a different approach, we can apply a regularization term $R(\cdot)$ in the optimization to implicitly restrict $z$ as follows:
\begin{equation} \label{recon2}
\hat{z} = \argmin\limits_{z\in \mathcal{Z}} \frac{1}{m} ||AG(z) - y||^2_2 + R(z)\,,
\end{equation}
where $R(z) = \lambda ||z||^2_2$ is a regularization term that imposes an isotropic Gaussian prior to the latent space and $\lambda$ is a balance hyper-parameter. It has been experimentally observed \citet{bora17} that this type of regularization helps with better exploration of the latent space of typical deep generative models and it can also stabilize the optimization process.

\subsection{Generative Adversarial Networks}

\begin{figure}
	\includegraphics[width=\columnwidth]{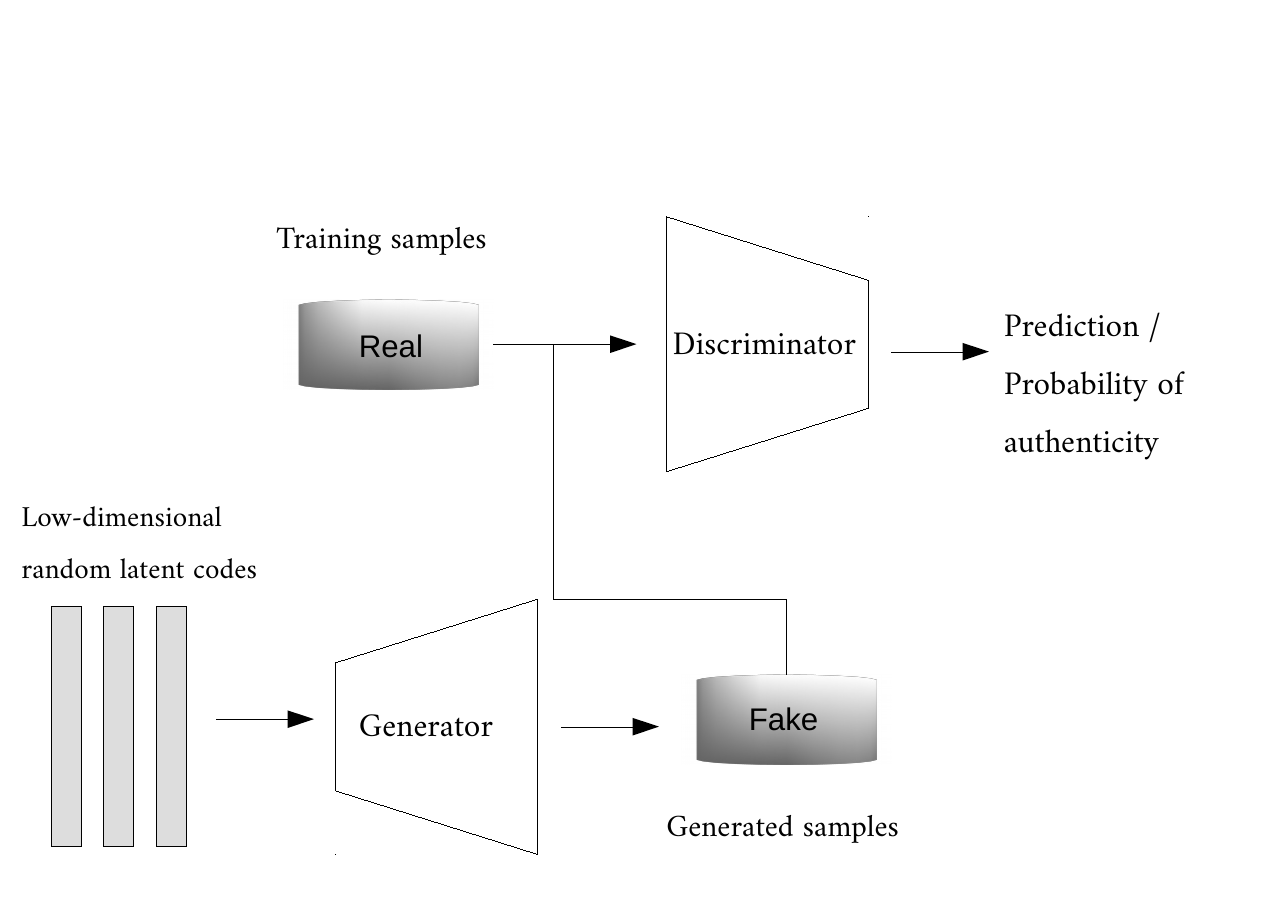}
	\caption{\label{figure:GAN}The training process in Generative Adversarial Networks accomplished through a (zero-sum) game that is played between two deep learning models: the generator and the discriminator. The generator tries to generate samples indistinguishable from  real ones, while the discriminator tries to distinguish between real samples and artificial samples produced by the generator.   }
\end{figure}

To obtain our deep generative prior, we first train a Generative Adversarial Network (GAN) (\citet{gans}). A Generative Adversarial Network is a generative model trained via an adversarial process, between two deep learning models the Generator and the Discriminator. On the one hand, the Generator works as a generative model: it takes as input random vectors sampled in the latent space (latent codes) and generates samples from the learned data distribution. On the other hand, the Discriminator takes as input either real samples from the training data or artificial samples produced by the Generator and works as a binary classifier predicting whether a given sample is real or artificial. Overall, the training procedure can be regarded as a two-player zero-sum game \citep{Nisan2007}, with the Generator trying to \enquote{fool} the Discriminator with artificially generated samples, while the Discriminator aims to efficiently distinguish between artificial and real samples. 
The procedure described above is depicted in Figure~\ref{figure:GAN}. It can be  formally described by the following optimization objective:
\begin{equation}\label{eq:GAN}
  \max\limits_{G}\min\limits_{D} \left( \mathbb{E}_{x \sim p_t} [1-D(x)] + \mathbb{E}_{z \sim p_z} [D(G(z))]  \right)\,,  
\end{equation}
 where $G$ is the Generator, $D$ is the Discriminator, $\mathbb{E}[\cdot]$ denotes the expected value,  $x \sim p_t$ denotes that $x$ is sampled from the training data distribution $p_t$, and $z \sim p_z$ denotes that $z$ is sampled from distribution $p_z$ over the latent space $\mathcal{Z}$. In \eqref{eq:GAN}, $\mathbb{E}_{z \sim p_z} [D(G(z))]$ is the probability that an artificially generated signal $G(z)$ produced by the Generator on latent code $z$ is classified as real by the Discriminator, while $\mathbb{E}_{x \sim p_t} [1-D(x)]$ is the probability that a real signal $x$ is classified as artificial by the Discriminator. According to the description above, the Generator aims to maximize their sum, while the Discriminator aims to minimize it. At equilibrium, both the Generator and the Discriminator behave optimally against each other. 

The motivating idea behind Generative Adversarial Networks comes from the difficulty of expressing a differentiable loss function that represents a complex data distribution. Instead, a trainable neural network, the Discriminator, is utilized to evaluate the generated samples. After the training procedure successfully converges to a (local) optimum, the Generator network can be used as a generative model, with random latent codes sampled from the latent space. 

In theory, GAN training should converge to an equilibrium of a zero-sum game. In practice, however, GAN training often exhibits difficulties due to instability, while it may happen that neither the Generator loss nor the Discriminator loss are informative of the stability of training \citet{gan_failures}. In other words, the training often does not converge and there is no principled way to choose a particular training iteration for which the Generator's performance is satisfactory. Instead, practitioners often choose such an iteration by manually inspecting random generated samples. Of course, this procedure can be accomplished for data such as images or sound, but not for complex data such as our spectra, where human evaluation is difficult, if even possible. 
Another important issue of GAN training is mode collapse \citet{goodfellow2016nips}, where the Generator collapses to a particular mode of the data distribution, namely, it keeps generating very similar samples, which are realistic but not diverse enough to cover the whole space of the actual data distribution. 

Several modifications and optimizations of GAN architecture and training have been proposed to tackle related issues \citet{dask18, pan2022unigan}. In the following paragraph we focus on the Wasserstein GAN (WGAN), a widely used variant of Generative Adversarial Networks.

\subsubsection{Wasserstein Generative Adversarial Networks}

The Wasserstein GAN is a GAN algorithm proposed in \citep{wasserstein} in order to tackle the issues described above. The most important modifications of this algorithm, which are the ones that we utilize in our experiments, are the following:

\begin{itemize}
    \item The Discriminator is no longer a binary classifier, deciding whether samples are real or artificial. Instead, it assigns a ``realness'' score to each sample. The idea is that the training of the Generator should be guided by the distance between the data distribution, empirically observed via the training data, and the Generator data distribution, empirically observed via the artificially generated samples. Now, the formulation of the loss function changes and the Discriminator loss is defined as the difference of the \enquote{realness} scores assigned to real samples and the scores assigned on artificially generated samples, whereas the Generator's score is defined as the \enquote{fakeness} score assigned to artificially generated samples.

    \item There are also modifications regarding  implementation details, which include more frequent training of the Discriminator compared to the Generator and weight clamping (i.e., weight value restriction) for the Discriminator. 
\end{itemize}

In our experiments we use the Wasserstein GAN model implementing the aforementioned modifications.

\subsection{Generative Latent Optimization}

\begin{figure}
	\includegraphics[width=\columnwidth]{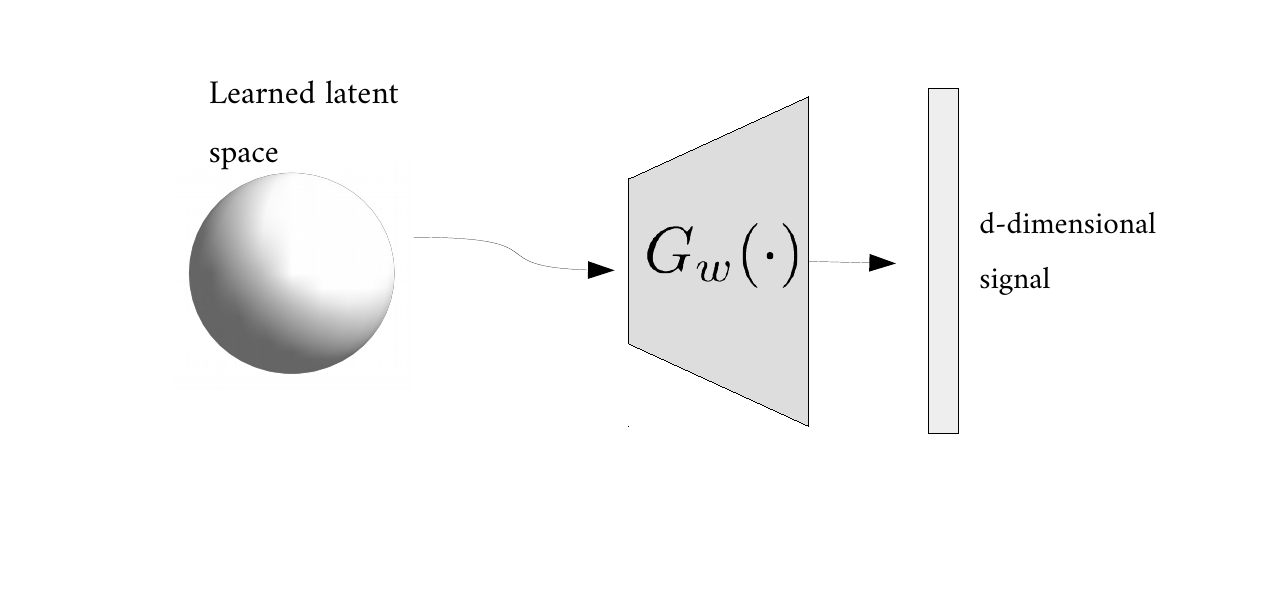}
	\caption{The training procedure of the GLO framework: 
The generator $G$ is trained to map learnable latent vectors to d-dimensional vectors, by minimizing a reconstruction loss. During the training phase, both the parameters of the generator and the tuning of the corresponding latent vectors are optimised.}
\end{figure}\label{fig:GLO}

To overcome issues that arise with GAN training, we also train a generative model using the Generative Latent Optimization (GLO) framework \citep{glo}, which allows us to train a relatively large generator network in order to facilitate generalization \citep{generalization_DNN}. The framework is based on the appropriate adaptation of the generator's latent space as well as its parameters, using a simple reconstruction loss. The generic framework of the GLO is depicted in Figure \ref{fig:GLO}. We use the GLO framework as an alternative to GANs. Unlike GLO, which consists of a simple loss minimization back-propagated to the latent space, GANs should ideally converge to an (approximate) equilibrium which is not guaranteed and/or requires excessive resources \citet{gan_equilibria}. Thus, when training GANs in practice it is common to examine the generated samples and stop the training when they are satisfactory. In the case of images this technique can be easily applied, but for SEDs this is not feasible since evaluating the fidelity of a SED is not a straightforward task. In fact, we use the ability to solve inverse problems with our trained generator as a proxy to evaluate whether training was successful. In particular, for testing purposes we use our trained generator and our reconstruction method with SEDs for which both linear measurements and the full signal are available. In this case, we can directly compute the reconstruction error using some error metric. Well trained generative models will yield low reconstruction error whereas poorly trained ones will not be successful at the reconstruction task. 

Let us examine the training procedure of a GLO model more closely. We train the generator $G \; : \; \mathcal{Z} \rightarrow \mathcal{X}$, where $\mathcal{Z}$ denotes the latent space and $\mathcal{X}$ the underlying class of SEDs which is described by the training set $\{ x_i \}_{i=1}^N$. Prior to training, we randomly initialize the latent codes $z_i \in \mathcal{Z}$ from a multi-dimensional normal distribution and pair them with each of the samples $x_i$. During training, the generator's weights and the latent codes $\{ z_i \}_{i=1}^N$ are jointly learned, as described by \eqref{glo_training}. The optimization is driven by a simple reconstruction loss $\mathcal{L}(\cdot)$, which in our case is Mean Squared Error (MSE). 

\begin{equation} \label{glo_training}
\min\limits_{G} \frac{1}{N} \sum_{i=1}^N \min\limits_{z_i \in \mathcal{Z}} \left[ \mathcal{L}(G(z_i), x_i) \right]
\end{equation}

More specifically, the gradient of the loss function with respect to the parameters of the generator and the latent code is back-propagated all the way through the network and to the latent space. This training procedure makes the latent space more structurally meaningful and suitable for reconstruction. To promote this feature, we project the latent codes onto the unit sphere during training, as in \citet{glo}.

\section{Experimental Setting}\label{experimental_setting}

\subsection{Data}

We prepare a library of simulated model spectra which we use to train our generative models and test our methods. The simulated model spectra are comprised of galaxies generated using the parameters listed in Table \ref{tab:data_library}. These models were generated using the CYGNUS radiative transfer models collection\footnote{The models are publicly available at \url{https://arc.euc.ac.cy/cygnus/}}  \citep{Efstathiou2022}.

The models used for simulating the spectra of galaxies have three components: the starburst galaxy models, the spheroidal host models and the AGN torus models, each with its own set of parameters, including optical depth, AGN torus opening angle, torus inclination, scaling parameters which determine the luminosities of each model category, etc. The emission for the starburst model is based on \citet{efstathiou00} and \citet{efstathiou09}, the emission of the spheroidal host model is based on \citet{Efstathiou2021}, and the AGN torus emission is based on \citet{efstathiou95} and \citet{efstathiou13}. 

For each model type, the parameters are randomized within the range of values displayed in Table~\ref{tab:data_library}, in order to create the library of simulated models used for training. The library consists of two sublibraries: the first contains $10,000$ complete spectra with $223$ points (wavelength bins) per spectrum which are roughly equally spaced logarithmically in the wavelength range 0.03-12,000 $\mu m$, whereas the second contains $10,000$ reduced spectra which are interpolated in $20$ wavelength bins as listed in Table 1. The age of the host galaxy is fixed to 1.32Gyr. The simulated spectra contain numerous emission features from PAHs and emission and absorption features from silicate dust but do not include optical emission lines. As we vary the scaling parameters of each model, we can simulate spectra of galaxies which are AGN-dominated, starburst-dominated, host-dominated or any combination of these. The models are normalized so they can apply to a galaxy of any luminosity, star formation rate or stellar mass. The simulated spectra therefore cover the parameter space that is appropriate for LIRGs, ULIRGs and HLIRGs.

We train two different kinds of generative models, one using the simulated spectra and another using the logarithmic values of the simulated spectra. This experimental choice is mainly motivated by the fact that the spectra have a very large dynamic range. The logarithmic spectra eliminate some of the high-frequency oscillations observed in the original spectra, which is proven to be beneficial for the training of our generative models. We discuss this effect extensively in the following sections.

\begin{table}
	\centering
	\begin{tabular}{|c|c|c|c|} 
		\hline
		Observed & Rest & & \\
            Wavelength & Wavelength & Telescope & Bands \\
		  $\lambda$ ($\mu m$) & $\lambda$ ($\mu m$) & & \\
		\hline

0.153  & 0.051  &  GALEX   & FUV \\   \hline
0.229  & 0.076  &  GALEX   & NUV \\    \hline
0.355  & 0.118  &  INT    &  u \\    \hline
0.65   & 0.217  &  INT   & r  \\  \hline
0.791  & 0.264  &  INT   &  i \\  \hline
0.908  & 0.303  &  INT   &    z  \\    \hline
0.978  & 0.326  & Subaru    &    Suprime N921 \\ \hline
1.241  & 0.414  & UKIRT    &    J  \\\hline
1.61   & 0.537  & UKIRT    &  H  \\       \hline
2.17   & 0.723  & UKIRT    &  K \\ \hline
3.556  & 1.185  & Spitzer     &  IRAC 1  \\     \hline  
4.501  & 1.500  & Spitzer     & IRAC 2  \\      \hline  
5.745  & 1.915  & Spitzer     & IRAC 3   \\     \hline 
7.918  & 2.639  & Spitzer     & IRAC 4 \\  \hline
23.594 & 7.865  & Spitzer     & MIPS24 \\ \hline
100    & 33.33  & Herschel     &  PACS \\      \hline  
160    & 53.33  & Herschel      & PACS \\    \hline
250    & 83.33  & Herschel     & SPIRE \\    \hline             
350    & 116.67  & Herschel     & SPIRE \\  \hline
500    & 166.66  & Herschel     & SPIRE \\ \hline

		\hline
	\end{tabular}
\caption{List of the photometry bands from the ultraviolet to the submillimetre used in this paper. They are representative of the bands available in a typical field studied by projects like the Herschel Extragalactic Legacy Project (HELP; \citet{Shirley2019,Shirley2021}). We give both the observed wavelength as well as the corresponding rest wavelength for galaxies at $z = 2$.}\label{tab:example_table}
\end{table}

\begin{table}
	\centering
	\begin{tabular}{|c|c|}
		\hline
		Simulated Spectra & Limited Photometry \\
		\hline
		223 wavelengths & 20 wavelengths\\
		range: 0.03-12,000 $\mu m$ & range: 0.05-170 $\mu m$\\
		\hline
	\end{tabular}
	\caption{Data statistics for $10,000$ samples.}
	\label{tab:data_library}
\end{table}

\subsection{Training Configuration}

For our experiments we train four generative models: Two Generative Latent Optimization models, one using the simulated spectra, namely GLO, and another using the logarithmic values of the simulated spectra, namely logGLO, and similarly, two Wasserstein GAN models, WGAN and logWGAN. 
The generators  are feed-forward neural networks with six hidden layers and leakyReLU activations (except for the output layer). For the GAN models, the discriminators are symmetric to the generators. 
For all our models, we use $90\%$ of the dataset as our training data and the remaining $10\%$ for testing.
Our complete spectra consist of measurements for $223$ wavelengths whereas our limited photometry measurements correspond to only $20$ wavelengths, as shown in Table~\ref{tab:data_library}. Theoretical and empirical results \citet{bora17} suggest that the latent vector size should be in the same order of magnitude as the number of limited measurements used for reconstruction ($20$ in our case). Thus, we choose $50$ dimensions for the latent space, which are sufficient for the representation and allow for efficient training and reconstruction. This choice of dimensionality allows a satisfactory level of compression for signals consisting of $223$ measurements that exhibit high-frequency oscillations.

For the Generative Latent Optimization training, we train our network for $10,000$ epochs with batches of $100$ spectra. We use Adam optimization \citep{adam} with learning rate $0.5$ for the network's parameters and $0.05$ for the latent codes, as well as 1d-batch normalization to accelerate the training procedure \citep{batchnorm}. We apply learning rate decay every $2,000$ epochs with decay rate $0.5$. We choose a simple Mean Squared Error (MSE) as our loss function and we also apply weight decay to avoid overfitting. 

For the Generative Adversarial Network training, we perform $15,000$ iterations of training with batches of $100$ spectra. We use RMSprop optimization, which is a suggested WGAN modification, with learning rate $0.001$ for both networks, as well as 1d-batch normalization. We train the Discriminator five times more often than the Generator. We apply learning rate decay every $3,000$ epochs with decay rate $0.5$. Finally, we clamp the Discriminator's weights between $[-0.01,0.01]$, as suggested by the WGAN algorithm.

\subsection{Reconstruction Procedure}

For the reconstruction, we limit the optimization procedure to $1,000$ epochs and choose a configuration similar to training, for each type of model.
For the Generative Latent Optimization models, we use projection onto the unit sphere, whereas for the Wasserstein GAN models, we use the regularization method with $\lambda = 10^{-5}$. In Table~\ref{quant}, we also evaluate the Generative Latent Optimization models with the regularization method for which we choose $\lambda = 10^{-3}$. We chose the learning rate for each trained model via a validation process. We use a learning rate of $0.001$ for the logGLO model, $0.01$ for the GLO model and $0.1$ for both WGAN models. Note that the Wasserstein GAN models do not converge, which is expected as discussed in the previous section. Thus, we choose among a small set of saved models along training via human evaluation of a random sample. We choose the model saved at the $15k$ iteration for WGAN and the model saved at the $10k$ iteration for logWGAN.

The project is developed using PyTorch \citep{pytorch}.

\begin{table}
\centering
\begin{tabular}{|c | c c|} 
 \hline
  & regularization & projection \\ [0.5ex] 
 \hline 
 GLO & 7.86 / 5 & 0.23 / 49  \\ 
 \hline
 logGLO & - & 0.19 / 72  \\ 
 \hline
 WGAN & 2.14 / 2 & -  \\ 
 \hline
 logWGAN & 6.1 / 0 & -  \\ [1ex] 
 \hline
\end{tabular}
\caption{Quantitative evaluation of each trained model and with each reconstruction measure. Average MSE / Number of spectra with $\chi^2 < 5$ for $100$ galaxies calculated in the wavelength range $5-35\mu m$.}
\label{quant}
\end{table}

\begin{figure}
    \centering
    \includegraphics[width=9cm,height=6cm]{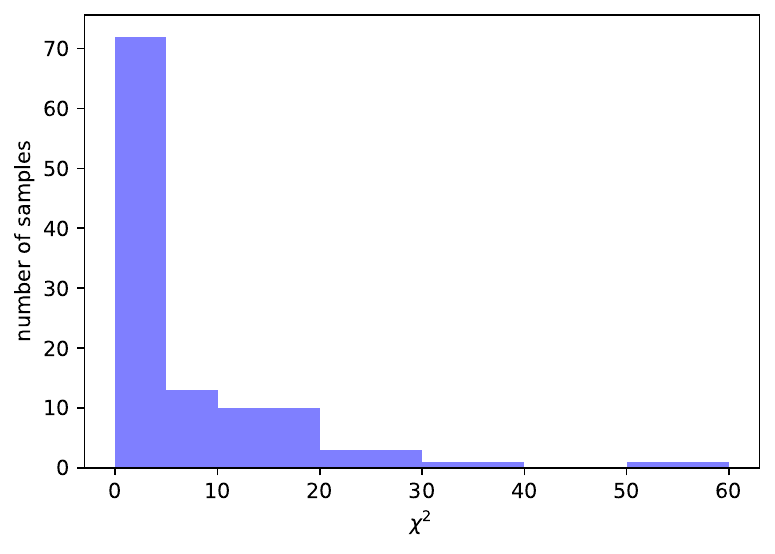}
    \caption{Quantitative results from the logGLO model. The histogram of the $\chi^2$ measure for $100$ samples selected at random is computed in the wavelength range $5-35\mu m$.}
    \label{fig:chi2_hist}
\end{figure}

\begin{figure}[tbh]
\centering
\includegraphics[width=0.8\columnwidth]{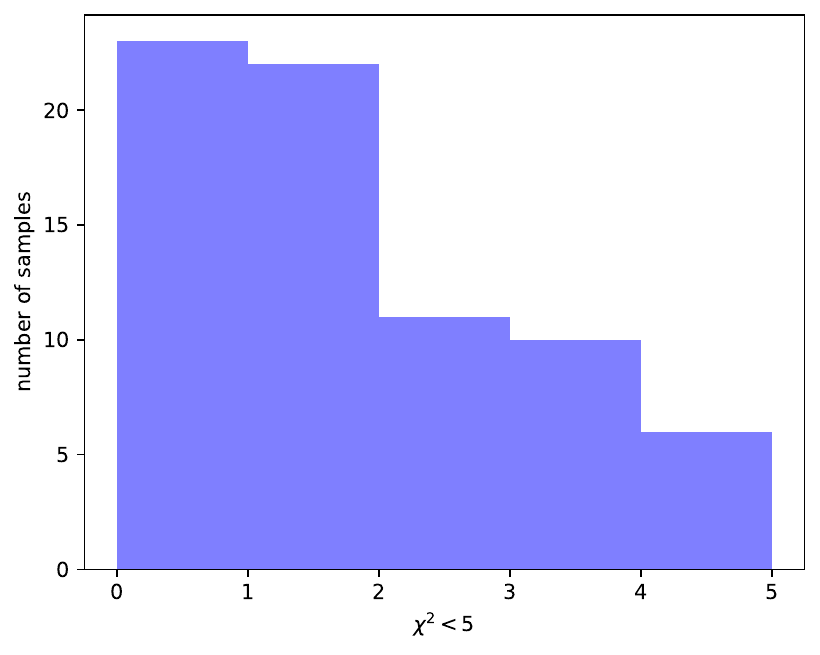}
\caption{Quantitative results from the logGLO model. The histogram of the $\chi^2$ measure for the $72$ among $100$ samples with $\chi^2 < 5$ selected at random is computed in the wavelength range $5-35\mu m$.}
\label{fig:chi2_hist_zoomed}
\end{figure}

\section{Results}\label{results}

\begin{figure*}[hbt!]
    \centering
    \includegraphics[width=0.7\linewidth]{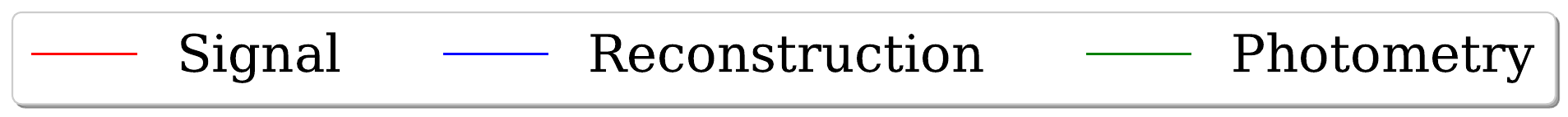}
    \includegraphics[width=\textwidth,height=4cm]{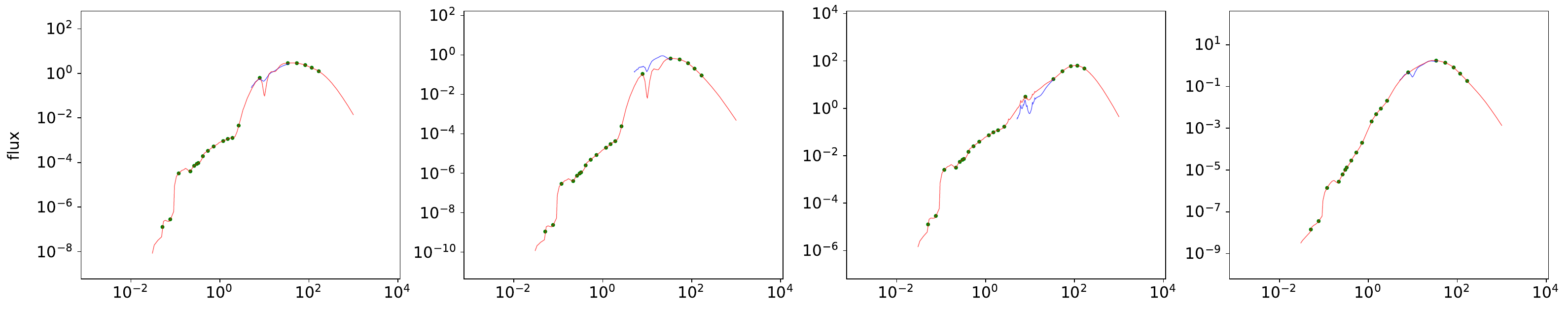}
    \includegraphics[width=\textwidth,height=4cm]{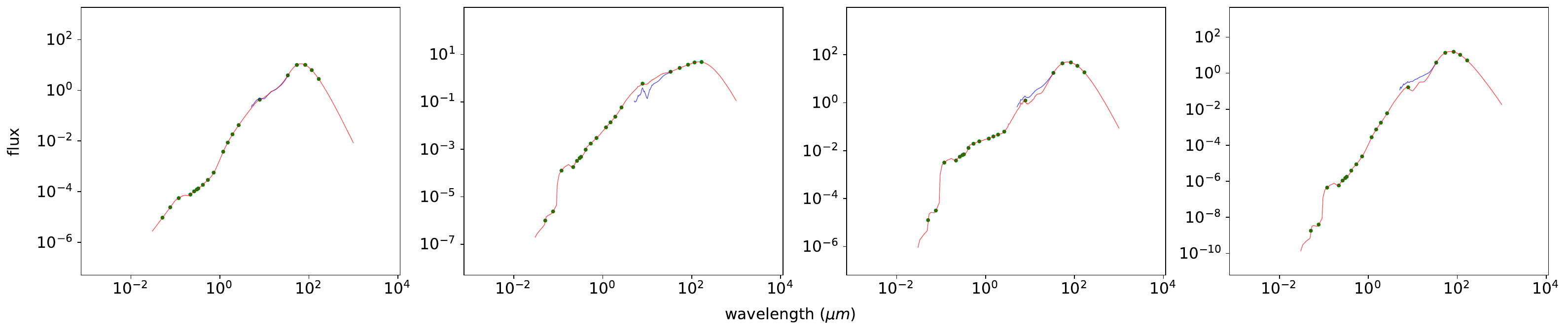}
    \caption{\textbf{GLO model}: Qualitative comparison between simulated spectra and reconstructed spectra using the GLO model for $8$ spectra randomly selected from the test set. The model receives the limited photometry measurements (green dots) and produces the reconstructed spectra (blue line) that should ideally be as close as possible to the simulated spectra (red line) which we treat as the ground truth. The reconstructed spectra are depicted in the region of interest ($5-35\mu m$).}
    \label{fig:glo_qual}
\end{figure*}

\begin{figure*}[hbt!]
    \centering
    \includegraphics[width=\textwidth,height=4cm]{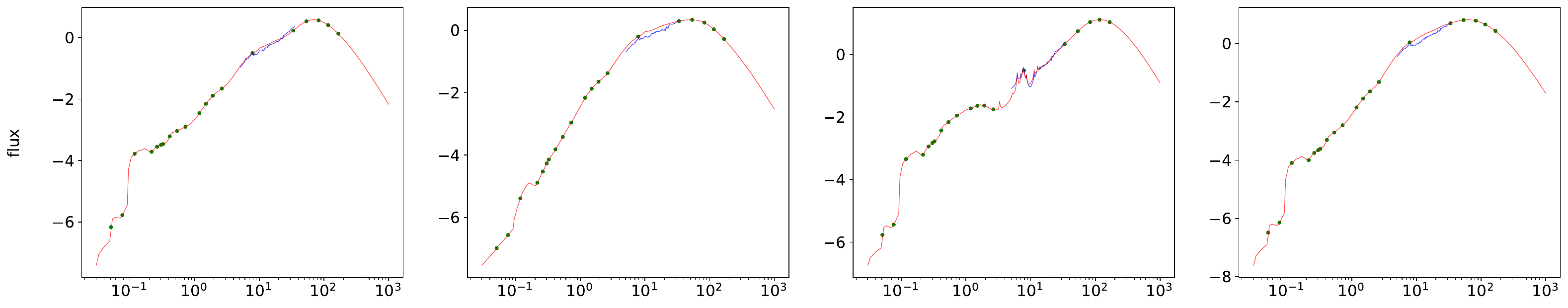}
    \includegraphics[width=\textwidth,height=4cm]{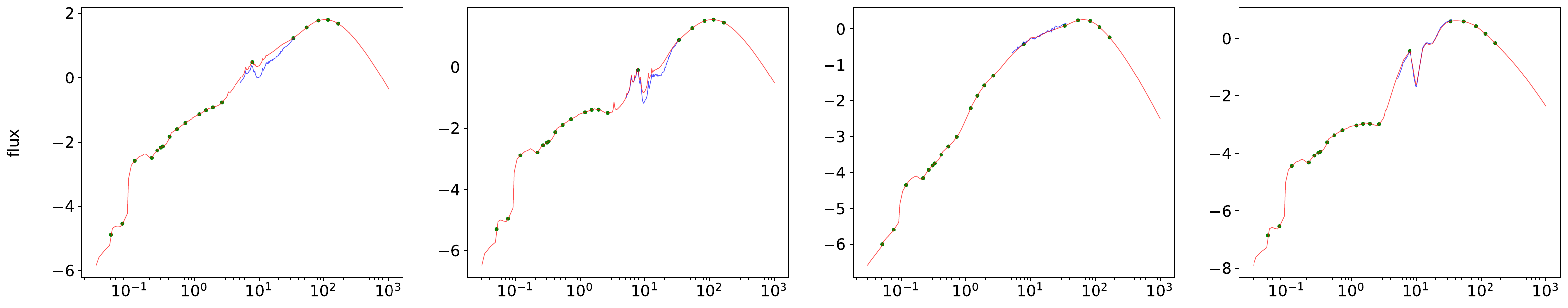}
    \includegraphics[width=\textwidth,height=4cm]{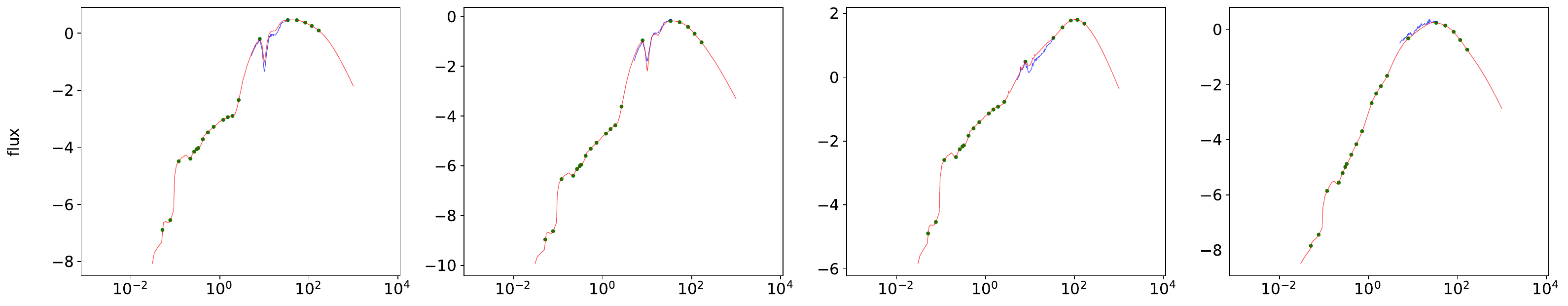}
    \includegraphics[width=\textwidth,height=4cm]{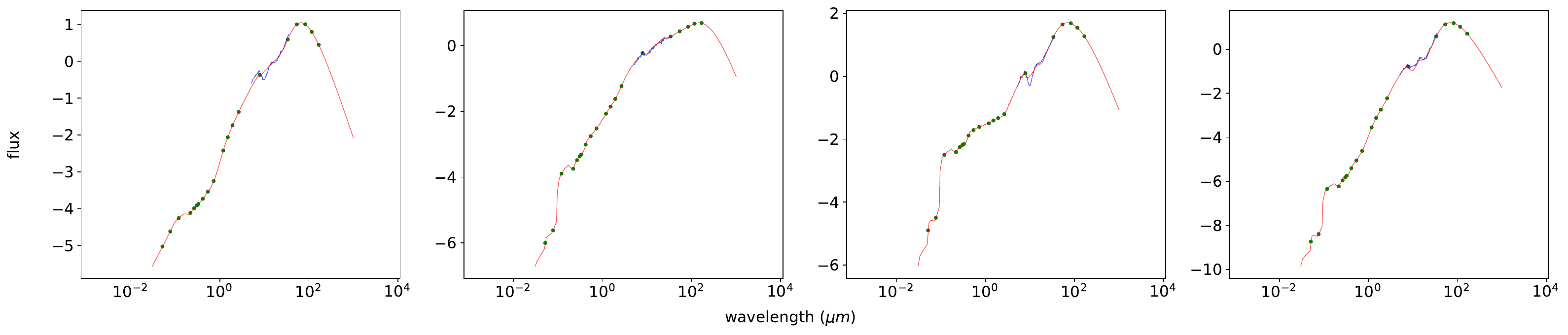}
    \caption{\textbf{logGLO model}: Qualitative comparison between simulated spectra and reconstructed spectra using the logGLO model for $8$ spectra randomly selected from the test set. The model receives the limited photometry measurements (green dots) and produces the reconstructed spectra (blue line) that should ideally be as close as possible to the simulated spectra (red line) which we treat as the ground truth. The reconstructed spectra are depicted in the region of interest ($5-35\mu m$).}
    \label{fig:logglo_qual}
\end{figure*}

We evaluate our approach, both qualitatively and quantitatively, for the different generative models that we trained. Our computational task is summarized as follows: For each sample galaxy from our test data, we consider the photometry data which are regarded as our incomplete measurements. We use our reconstruction method outlined above to estimate a reconstruction of the 5-35$\mu m$ spectrum of a particular galaxy. We focus our evaluation on the $5-35 \mu m$ wavelength range mostly because this wavelength range is rich in features providing information about star formation and AGN activity in the galaxies. The method can, of course, in principle be applied to reconstruct any other part of the spectrum. We evaluate our reconstruction with respect to the simulated spectrum of the galaxy from our curated dataset, namely our signal. 

\begin{figure*}[hbt!]
    \centering
    \includegraphics[width=\textwidth,height=4cm]{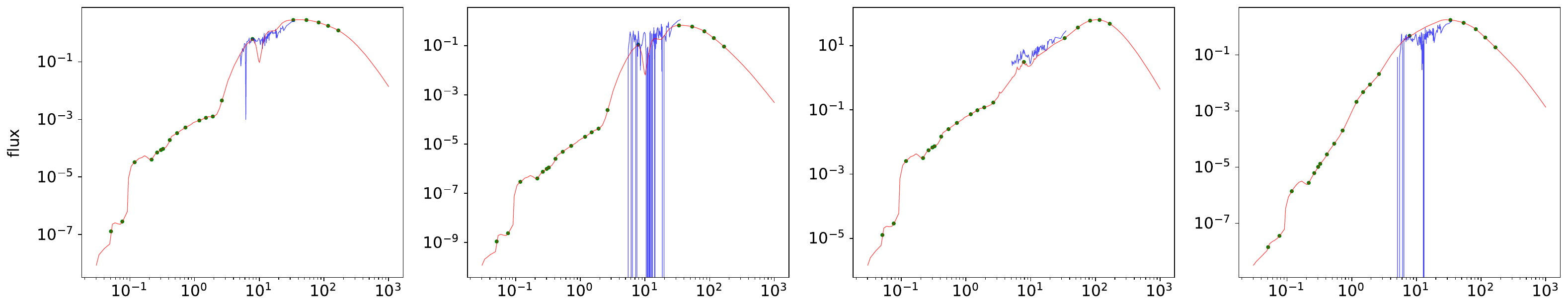}
    \includegraphics[width=\textwidth,height=4cm]{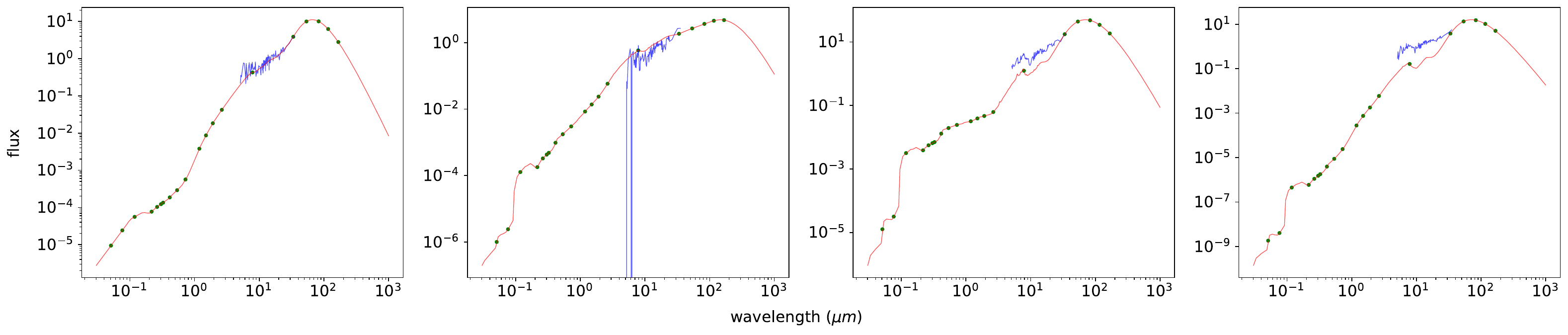}
    \caption{\textbf{WGAN model}: Qualitative comparison between simulated spectra and reconstructed spectra using the WGAN model for $8$ spectra randomly selected from the test set. The model receives the limited photometry measurements (green dots) and produces the reconstructed spectra (blue line) that should ideally be as close as possible to the simulated spectra (red line) which we treat as the ground truth. The reconstructed spectra are depicted in the region of interest ($5-35\mu m$).}
    \label{fig:wgan_qual}
\end{figure*}

\begin{figure*}[hbt!]
    \centering
    \includegraphics[width=\textwidth,height=4cm]{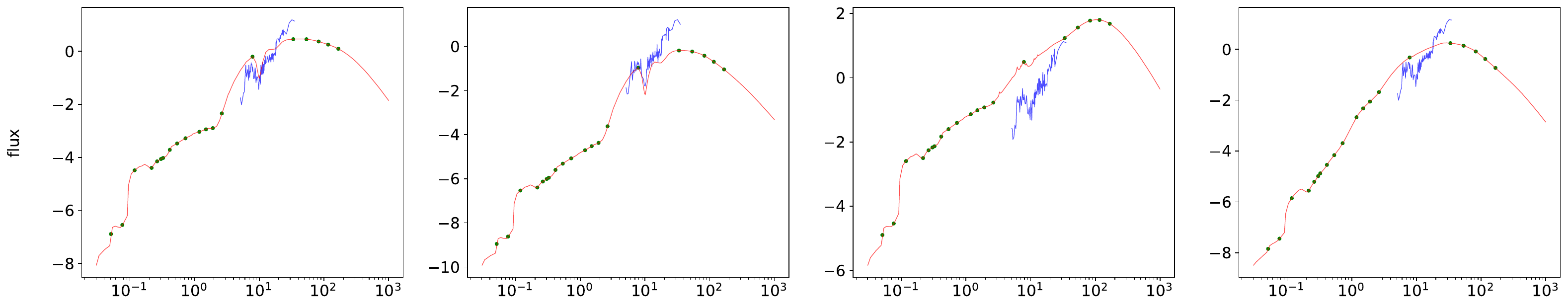}
    \includegraphics[width=\textwidth,height=4cm]{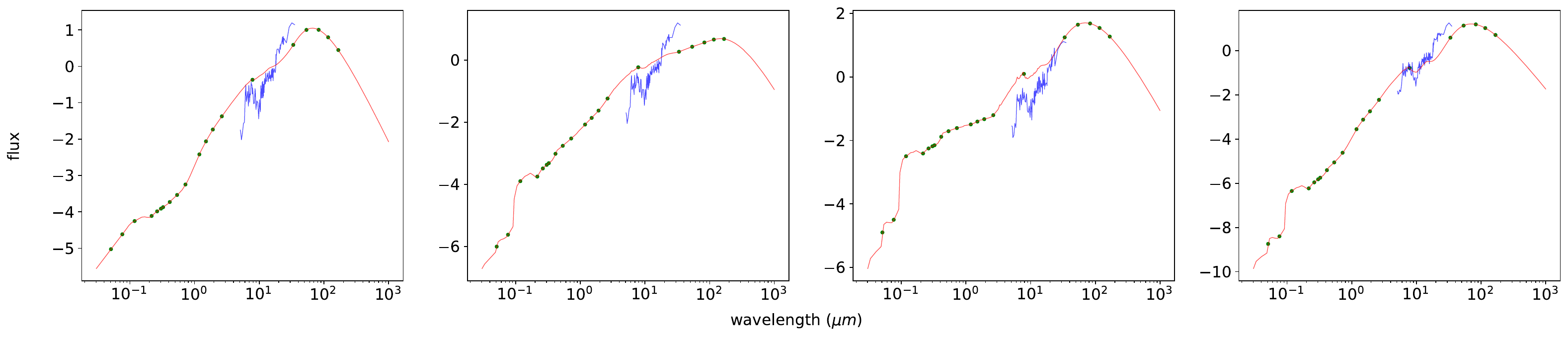}
    \caption{\textbf{logWGAN model}: Qualitative comparison between simulated spectra and reconstructed spectra using the logWGAN model for $8$ spectra randomly selected from the test set. The model receives the limited photometry measurements (green dots) and produces the reconstructed spectra (blue line) that should ideally be as close as possible to the simulated spectra (red line) which we treat as the ground truth. The reconstructed spectra are depicted in the region of interest ($5-35\mu m$).}
    \label{fig:logwgan_qual}
\end{figure*}

For the qualitative evaluation, we randomly select eight galaxies from our test set and reconstruct their full spectrum based on the corresponding limited photometry measurements. In Figures \ref{fig:glo_qual}, \ref{fig:logglo_qual}, \ref{fig:wgan_qual} and \ref{fig:logwgan_qual} we plot our reconstructions in the important region of $5-35 \mu m$ for our four trained generative models, along with the corresponding underlying signal and limited photometry measurements. We plot the spectra as well as their reconstructions in logarithmic scale for all of our models, because of the large dynamic range of the spectra. 

\begin{figure*}[hbt!]
    \centering
    \includegraphics[width=\textwidth,height=4cm]{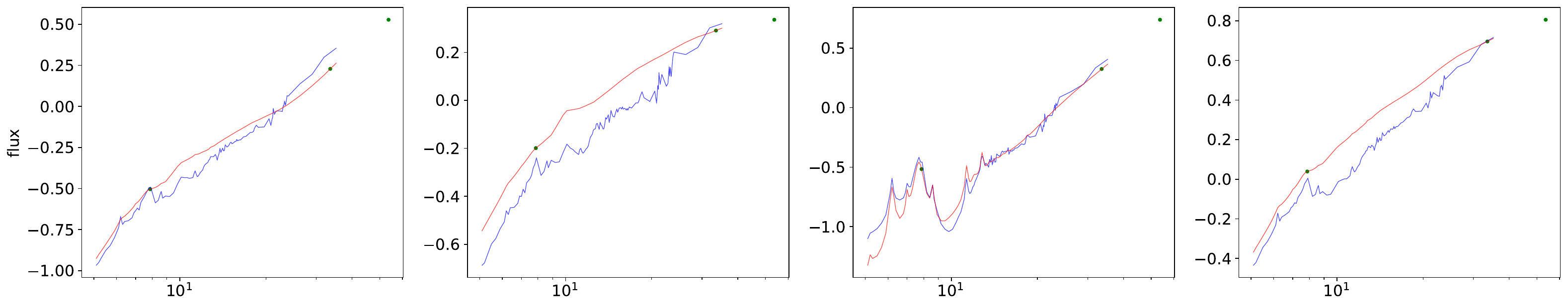}
    \includegraphics[width=\textwidth,height=4cm]{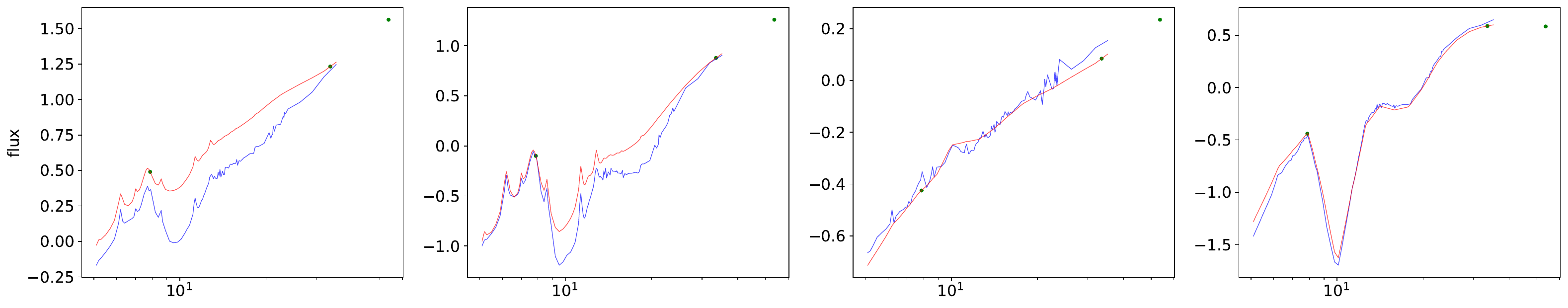}
    \includegraphics[width=\textwidth,height=4cm]{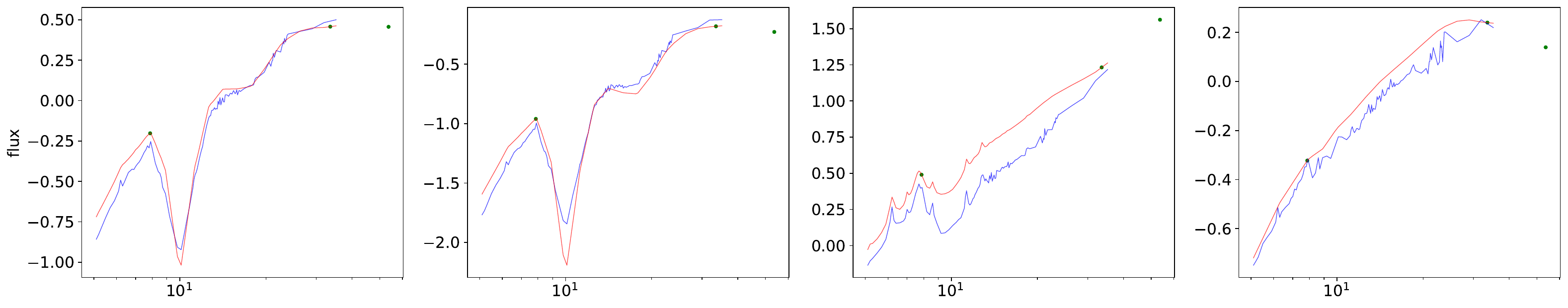}
    \includegraphics[width=\textwidth,height=4cm]{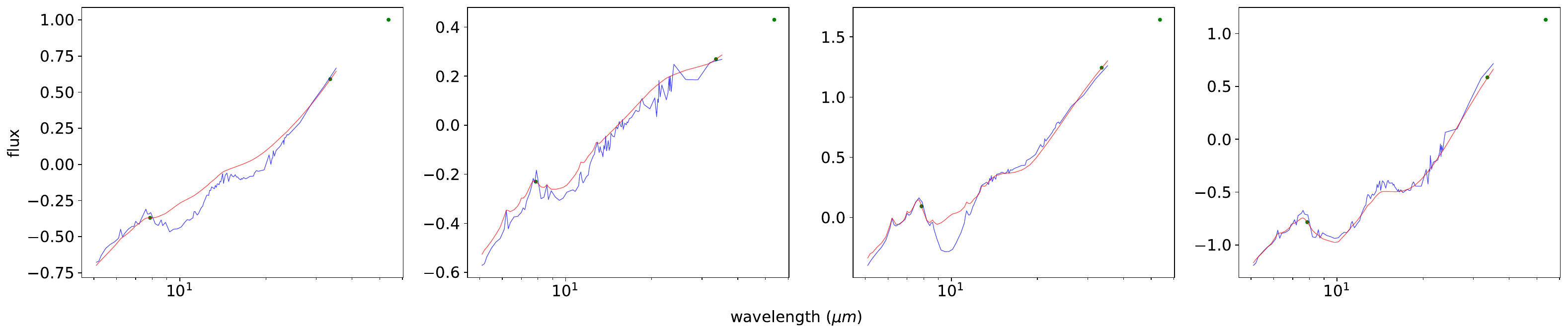}
    \caption{\textbf{logGLO model, focused}: Qualitative comparison between simulated spectra and reconstructed spectra using the logGLO model for $16$ spectra randomly selected from the test set. The model receives the limited photometry measurements (green dots) and produces the reconstructed spectra (blue line) that should ideally be as close as possible to the simulated spectra (red line) which we treat as the ground truth. The reconstructed spectra are depicted in the region of interest ($5-35\mu m$). Here we focus on the mid-infrared part of the spectrum which is our main interest in this study.}
    \label{fig:logglo_qual_focus}
\end{figure*}

We observe that our GLO model (Figure \ref{fig:glo_qual}) mostly follows the shape of the underlying spectrum, although it introduces some spikes which indicate that the high-frequency oscillations observed in the simulated spectra may negatively affect the training and the results in the reconstruction. Our best performing model is the logGLO model (Figures \ref{fig:logglo_qual} and \ref{fig:logglo_qual_focus}), which very closely follows the shape of the original signal and in most cases correctly predicts the important spikes. This indicates that training with logarithmic spectra allows for the model to focus on large spikes and disregard noisy oscillations. 
For our GAN models, we observe much worse performance, which indicates that for both models the adversarial training did not converge to a satisfactory domain representation. The WGAN model displays certain artifacts, because it predicts very small (close to zero) values, whereas the logWGAN does not have this feature. This observation again shows the benefit of training with logarithmic spectra. 

In Table \ref{quant} we examine quantitatively the performance of our models. For each configuration, we compute: \begin{enumerate}
    \item The reconstruction MSE and
    \item The $\chi^2$ statistic
\end{enumerate}  
of $100$ randomly selected spectra from our test set. We define the $\chi^2$ as follows:

Suppose we reconstruct the mid-infared (5-35$\mu m$) spectra $S_{rec}$ at $N_{mir}$ points at which we also know the model spectra $S_{mod}$. We then compute the $\chi^2$ for each galaxy as a measure of how close the reconstruction matches the model spectra, that is:  

\begin{equation}\label{eq:chisq}
{\chi^2} =  \frac{1}{N_{mir}} \sum_{i=1}^{N_{mir}}  \frac{(S_{rec}[i] - S_{mod}[i])^2}{(0.1 \times S_{rec}[i])^2 + (0.1 \times S_{mod}[i])^2}\,, \end{equation}
where $S_{rec}[i]$ (resp. $S_{mod}[i]$) denotes the value at the $i$-th point of the reconstructed (resp. model) spectra $S_{rec}$ (resp. $S_{mod}$). 
In the denominator of Equation \eqref{eq:chisq}, we add in quadrature our estimated uncertainty in the model and the reconstruction, 
assuming they are both 10\%. 
Our definition of $\chi^2$ is motivated by Equation (2) of \citet{lanz14} who compared simulated spectra of galaxies with data. \citet{lanz14} assigned 30\% uncertainty to simulated spectra and 10\% uncertainty to data.   

Usually in this kind of analysis, if the model agrees with the data we expect $\chi^2$ to be $\sim 1$,
yet a relatively small $\chi^2$ is considered satisfactory.
For the qualitative analysis that follows, 
we assume that $\chi^2 < 5$ indicates a good reconstruction.
We further justify this by showing that $\chi^2 < 5$ implies that the reconstructed spectra approximates well the model spectra, i.e., by a small multiplicative factor. First, we let $S_{req}[i] = \alpha S_{mod}[i]$, for some $\alpha \geq 0$ (i.e., the $i$-th value $S_{req}[i]$ of the reconstructed spectra approximates the $i$-th value $S_{mod}[i]$ of the model spectra multiplicatively within a factor of $\alpha$). Then, the $i$-th term of the summation becomes $\frac{100(1-\alpha)^2}{1+\alpha^2}$, which is less than $5$, only if $\alpha \in [0.72395, 1.38132]$, and raises fast as $\alpha$ either decreases (i.e., $\frac{100(1-\alpha)^2}{1+\alpha^2} \approx 11.8$, for $\alpha = 0.6$) or increases (i.e., $\frac{100(1-\alpha)^2}{1+\alpha^2} \approx 10.1$, for $\alpha = 1.6$) outside that interval. Therefore, a $\chi^2$ value less than $5$ in Equation \eqref{eq:chisq} implies that the reconstructed spectra approximates the model spectra by a multiplicative factor in (or quite close to) the range $[3/4, 4/3]$ for the vast majority of  points. Additionally, for $\chi^2 < 1$, the approximation range is roughly $[7/8, 8/7]$, while for $\chi^2 < 3$, the approximation range is $[0.78, 1.28]$.

%

Our quantitative comparison shows that 72\% of the galaxies have a $\chi^2 < 5$ in the case of the logGLO model (Figure \ref{fig:chi2_hist}). In Figure~\ref{fig:chi2_hist_zoomed} we present a histogram distribution of the $\chi^2$ measure for these 72 galaxies, using the logGLO model. We observe that about $1/3$ of these 72 galaxies have $\chi^2$ value less than $1$ and another $1/3$ have $\chi^2$ value between $1$ and $2$. The remaining $1/3$ are distributed between $2 < \chi^2 < 5$. Considering the total sample of 100 galaxies, we achieve successful reconstructions in about 60\% of the galaxies with $\chi^2 < 3$ and in about 70\% of the galaxies with $\chi^2 < 4$.
This shows that our qualitative observations are quantitatively supported. Both our GLO models have lower errors compared to our GAN models. In term of the reconstruction methods, we observe that the regularization benefits the GAN models possibly due to the flexibility it offers. On the other hand, the GLO models largely benefit by the projection method, which is expected since they incorporate projection during training in order to enforce structure in their latent space.


\begin{table}
\centering
\footnotesize
\begin{tabular}{|c || c c c c|} 
 \hline
 Latent dim & 20 & 30 & 40 & 50 \\ [0.5ex] 
 \hline
 logGLO + proj & 0.38 / 53 & 0.32 / 69 & 0.2 / 64 & 0.19 / 72 \\ 
 \hline
\end{tabular}
\caption{Impact of latent space dimensionality on performance. The results are for our best model (logGLO) and reconstruction with projection. Performance is measured with two metrics: Average MSE (AMSE) and Number of spectra with $\chi^2 < 5$ for $100$ galaxies calculated in the wavelength range $5-35\mu m$.}
\label{quant_latent}
\end{table}

Finally, for our best model and reconstruction method we explore the effect of latent space dimensionality on performance. Specifically, in Table \ref{quant_latent} we examine quantitatively the performance of logGLO and reconstruction with projection for different choices of the latent space dimensionality, i.e., $20$, $30$, $40$ and $50$. We observe that the average MSE, which is a continuous measure, drops significantly from $30$ to $40$ but stabilizes thereafter. The number of spectra with $\chi^2 < 5$ displays more variance because of its discrete nature and the fact that it is computed on a random subset of $100$ galaxies, but also seems to converge at around $50$ dimensions.

\section{Discussion}\label{discussion}

Our analysis shows that with the logGLO model we can reconstruct very well the mid-infrared spectra of galaxies in $\sim$ 70\% of the cases. The model successfully reconstructs for example the spectra of galaxies where a deep silicate absorption feature is observed around $9.7\mu m$, which is indicative of galaxies like the ULIRG IRAS~08572+3915 whose SED is dominated by emission from an AGN torus which is viewed almost edge-on \citep{Efstathiou2014,Efstathiou2022}. It also gives a good prediction of the presence of PAH emission features in the spectrum which are considered to give a good indication of enhanced star formation activity. From the results presented in Figure \ref{fig:logglo_qual_focus} it appears that it is easier to reconstruct the spectra when they are dominated by a silicate absorption feature instead of PAH features.

The successful reconstruction of the mid-infrared spectra of LIRGs and ULIRGs by the logGLO model implementation is particularly useful for these types of galaxy, as it can retrieve important information regarding the emission mechanisms of these galaxies based on the aforementioned physical features with a high degree of accuracy. This improved ability of the logGLO model compared to the rest of the deep generative models considered in this work is showcased in Figure \ref{fig:logglo_qual_focus} where we focus on the mid-infrared spectrum which is our main interest in this work. 

Although the analysis presented here cannot substitute, of course completely the actual observations of galaxies in the mid-infrared part of the spectrum, it can be very helpful in giving a good indication of the spectra of galaxies in the rest frame wavelength range between 5-35$\mu m$ which is likely to remain inaccessible for galaxies at redshift $z > 2$ in the foreseeable future. A prime example of this limitation is JWST, whose Mid-Infrared Instrument (MIRI) can observe  up to $28.0 \mu m$ and can therefore cover the wavelength range of interest for $z < 2$. 
The method can also be used for example to construct diagnostic diagrams like the \textit{fork} diagram, introduced by \citet{Spoon2007} and other diagnostic diagrams, for samples of galaxies at various redshifts. It can also be very useful in the planning of observing strategies with future telescopes.

\subsection{Limitations and Future Work}

Despite the illustrated potential of our current methodology, it should be considered a first step towards harnessing the power of deep generative models for spectra reconstruction from limited measurements. We emphasize that our work heavily relies on simulated spectra for training and evaluation. Future research should directly incorporate observed spectra and ideally combine all available data sources. In this pursuit, particular attention must be devoted to addressing observational effects inherent in real-world data, such as redshift variance, measurement errors, and variations in filters employed across different surveys. Instead, we purposefully focused on a curated dataset without specifically considering the role of these observational effects, thus essentially serving as a proof-of-concept for our proposed methodology.

The success of the method used in this study can motivate further studies in future work.  We plan for example to explore the success of the method in reconstructing spectra at different redshifts as the coincidence of important features like the PAH features and silicate features with key bands like the 24 and 70$\mu m$ bands may prove critical. We also plan to use the method to reconstruct the spectra of galaxies which already have observed mid-infrared spectra with NASA's Spitzer Space Telescope (e.g. the HERUS galaxies, \citet{Farrah2013}) or JWST using only the photometry and compare with the observed spectra. Comparison of reconstructed with real data will help to identify potential problems and limitations in the reconstruction method but also in the models used for simulating the training data. It would also be of interest to explore the impact of using different combinations of models for generating the library of simulated models and possibly expanding the size of the libraries to improve the performance of the method.
On the machine learning side, future work includes devoting more computational resources to explore the effect of larger and more powerful models in parallel to increasing the amount of training data (e.g., in our case using $10 \times$ more simulated spectra). Recent advances in other fields, such as Computer Vision (CV) \citet{krizhevsky2012imagenet,he2016deep} and Natural Language Processing (NLP) \citet{devlin2018bert,vaswani2017attention}, suggest that our method would also benefit from an increase in scale of both training data and model parameters.

\section{Conclusions}\label{conclusions}

We have developed and tested two different kinds of deep generative models for reconstructing the mid-infrared spectra of galaxies from limited photometry that spans the optical to submillimeter range. The first generative model tested is an advanced one, which  utilizes game theory for the training of the generative model,  i.e. a GAN, and the second one is a simpler one which utilizes simple reconstruction losses for the training, i.e. a GLO generative model. 
We found that the GLO model, and in particular a variation of it that uses the logarithm of the initial input signals, i.e. the logGLO model outperforms the GANs model, managing to successfully reconstruct the spectra predicted from simulated spectra produced by radiative transfer models.
This is attributed to the use of a relatively small set of training data of the experimental setting investigated, which makes the GLO more successful compared to a GAN, for which its success relies heavily on the availability of a sufficiently large training set.  

Perhaps the most significant contribution of this work is that it provides a framework and a case study for the further exploration of various deep generative models for the reconstruction of signals in the fields of astrophysics and cosmology, which are not limited to astrophysical images but extend to other sources of information, such as spectral energy distributions. 

\section*{Acknowledgements}

The authors acknowledge support from the project EXCELLENCE/1216/0207/ GRATOS funded by the Cyprus Research \& Innovation Foundation. Most of this work was carried out while AR was a PhD student at National Technical University of Athens and partly funded by the GRATOS project. AE and VPL acknowledge support from the project CYGNUS funded by the European Space Agency.

\section*{Data Availability Statement}

The data underlying this article are available in the article or are publicly available in databases like The Cornell Atlas of Spitzer/Infrared Spectrograph Sources (CASSIS) and the NASA Extragalactic Database (NED).

\cleardoublepage

\bibliographystyle{mnras}
\bibliography{bibliography}

\appendix

\section{Description of the CYGNUS models used for generating the simulated galaxy data}

\renewcommand{\thetable}{A.\arabic{table}}
\setcounter{table}{0}
\begin{table*}[!htbp]
	\centering
\caption{Parameters of the CYGNUS models used to produce the simulated spectra of galaxies, symbols used, their assumed ranges and summary of other information about the models. There are 3 additional scaling parameters for the starburst, AGN and spheroidal models, $f_{SB}$,  $f_{AGN}$ and $f_s$ respectively.}
	\label{tab:example_table}
	\resizebox{\textwidth}{!}{\begin{tabular}{llll} 
		\hline
		Parameter &  Symbol & Range &  Comments\\
		\hline
                 &  &  & \\
{\bf Starburst}  &  &  & \\
                 &  &  \\
Initial optical depth of giant molecular clouds & $\tau_V$  &  50-250  &  \citet{efstathiou00}, \citet{efstathiou09} \\
Starburst star formation rate e-folding time       & $\tau_{*}$  & 10-30Myr  & Incorporates \citet{BruzualCharlot1993,BruzualCharlot2003}  \\
Starburst age      & $t_{*}$   &  5-35Myr &  metallicity=solar, Salpeter Initial Mass Function (IMF) \\
                  &            &  & Standard galactic dust mixture with PAHs \\
                   &           &         &    \\
{\bf Spheroidal Host}  &  &  &  \\
                 &  &  \\
Spheroidal star formation rate e-folding time      & $\tau^s$  &  0.125-8Gyr  & \citet{Efstathiou2003}, \citet{Efstathiou2021}  \\
Starlight intensity      & $\psi^s$ &  1-17 &  Incorporates \citet{BruzualCharlot1993,BruzualCharlot2003} \\ 
Optical depth     & $\tau_{v}^s$ & 0.1-15 &  metallicity=40\% of solar, Salpeter IMF\\ 
                  &            &  & Standard galactic dust mixture with PAHs \\
                  &            &  &  \\
{\bf AGN torus}  &  &    &  \\
                 &  &  &  \\
Torus equatorial UV optical depth   & $\tau_{uv}$  &  250-1450 &  Smooth tapered discs\\  
Torus ratio of outer to inner radius & $r_2/r_1$ &  20-100 & \citet{efstathiou95}, \citet{efstathiou13} \\   
Torus half-opening angle  & $\theta_o$  &  30-75\degree & Standard galactic dust mixture without PAHs\\ 
Torus inclination     & $\theta_i$  &  0-90\degree &  \\ 
                 &            & \\
	\end{tabular}}
\end{table*}


\bsp	
\label{lastpage}
\end{document}